\documentclass[aps,pra,showpacs,twoside,twocolumn,10pt]{revtex4-2}
\usepackage[utf8]{inputenc}
\usepackage[sc,osf]{mathpazo}
\usepackage{amsmath}
\usepackage{latexsym}
\usepackage{amssymb}
\usepackage{amsfonts}
\usepackage[colorlinks=true,citecolor=blue,urlcolor=blue]{hyperref}
\usepackage{color}
\usepackage{graphics,epstopdf}
\usepackage{subcaption}
\usepackage{soul}
\usepackage{graphicx}
\usepackage{capt-of}
\usepackage{lipsum}
\usepackage{adjustbox}
\usepackage[normalem]{ulem}
\usepackage[table,xcdraw]{xcolor}
\usepackage{braket}
\usepackage{physics}
\usepackage{mathtools}

\usepackage{comment}
\usepackage{ragged2e}
\usepackage[font=small,labelfont=bf,justification=justified,singlelinecheck=false]{caption}
\usepackage{bm}
\newcommand{\avg}[1]{\langle #1 \rangle}
\newcommand{\id}{\mathbf{I}}
\newcommand{\bX}{\bm{X}}
\newcommand{\bG}{\bm{G}}
\newcommand{\bA}{\bm{A}}
\newcommand{\bD}{\bm{D}}
\newcommand{\bB}{\bm{B}}
\newcommand{\bOmega}{\bm{\Omega}}
\newcommand{\bsigma}{\bm{\sigma}}
\newcommand{\bSigma}{\bm{\Sigma}}
\definecolor{indiagreen}{rgb}{0.07, 0.53, 0.03}
\definecolor{teal}{rgb}{0.0, 0.53, 0.53}

\begin{document}
\title{
%\adi{Squeezing and memory enhanced frequency estimation in noisy continuous-variable systems}\\
Beating noise in frequency estimation with squeezing and  memory\\ in  continuous-variable systems
%\adi{Metrological gain with squeezing and  memory in  noisy continuous-variable systems}
}
%\title{Quantum Fisher information for the parameter $\omega$, with special cases}
\author{Ayan Patra$^{1,2}$, Manju$^{1,2}$, Aditi Sen(De)$^{1,2}$, Matteo G. A. Paris$^{3,4}$}

\affiliation{\(^1\)Harish-Chandra Research Institute, Chhatnag Road, Jhunsi, Allahabad - 211019, India\\
\(^2\) Homi Bhabha National Institute, Training School Complex, Anushakti Nagar, Mumbai 400 094, India}
\affiliation{$^3$ Dipartimento di Fisica “Aldo Pontremoli”, Università di Milano, I-20133 Milan, Italy}
\affiliation{$^4$ INFN, Sezione di Milano, I-20133 Milan, Italy}
\date{\today}

\begin{abstract}{
    Quantum metrology promises precision beyond classical limits, yet environmental noise typically degrades the quantum resources required for such enhancement. In this work, we investigate frequency estimation in noisy continuous-variable systems, focusing on two complementary strategies to mitigate decoherence: Hamiltonian engineering and the exploitation of non-Markovian dynamics. By embedding squeezing directly into the system Hamiltonian, we show that the quantum Fisher information (QFI) may acquire a tunable higher-order time dependence, leading to enhanced sensitivity in the short-time regime. Moving beyond the Markovian approximation, we employ the quantum Brownian motion model to demonstrate that structured environments with finite memory can induce information backflow, temporarily restoring and even improving estimation precision relative to the unitary limit. We further assess the achievability of these bounds via Gaussian measurements, identifying regimes where homodyne, heterodyne, and optimized general-dyne measurements saturate the QFI, and noting that stronger squeezing widens the gap, potentially requiring non-Gaussian measurement strategies. Our results establish that jointly tailoring system Hamiltonian and environmental memory offers a viable route toward robust quantum-enhanced frequency estimation in open systems.}
\end{abstract}

\maketitle

\section{Introduction}
Quantum metrology exploits quantum features to enhance the precision of parameter estimation, surpassing classical limits \cite{giovannetti2004quantum,giovannetti2006quantum}. It underlies a wide range of applications, including super-resoultion imaging \cite{kose2023superresolution}, ultra-precise clocks \cite{rosenband2008frequency}, gravitational-wave detection with the laser interferometer gravitational-wave observatory (LIGO) \cite{ligo2011gravitational,oelker2016audio}, ultra-sensitive magnetometry \cite{budker2007optical,barry2020sensitivity,patel2020subnanotesla}, gravimetry \cite{kanno2021indirect,stray2022quantum,seveso2017quantum}, thermometry \cite{weng2014nano,brunelli2012qubit,paris2016achieving}, and dark matter searches \cite{backes2021quantum,hochberg2022new}. By harnessing quantum resources like entanglement, coherence and squeezing \cite{nagata2007beating,pezze2009entanglement,unternahrer2018super,zhang2025entanglement,caves1981quantum,wineland1992spin}, it is possible to overcome the shot-noise limit (SNL) and, in the ideal scenarios, to approach the Heisenberg limit (HL). These features make quantum metrology one of the most promising directions in quantum technologies \cite{paris2009quantum,toth2014quantum,schnabel2017squeezed}.

%In realistic scenarios, 
In practice, however, no quantum probe is perfectly isolated, and the interaction between the quantum probe and its surroundings is inevitable, leading to decoherence and dissipation \cite{breuer2002theory}. Such a situation typically degrades the quantum resources responsible for enhanced sensitivity \cite{alipour2014quantum,smirne2016ultimate,haase2018fundamental,tamascelli2020quantum,monras2007optimal}. Within the commonly adopted Markovian description, the environment carries no memory, so information leaked from the probe is irretrievably lost. As a consequence, the quantum Fisher information (QFI) generally reaches a maximum at a finite interrogation time and subsequently declines as noise accumulates~\cite{krischek2011useful,demkowicz2014using,wang2018entanglement,falaye2017investigating,huang2016usefulness,saleem2023optimal}. 
A qualitatively different scenario emerges when the reservoir has finite memory and cannot be treated as memoryless. The resulting non-Markovian evolution depends on the prior history of the system and can exhibit temporary reversals of information flow from the environment back to the probe. Such backflow may revive coherence, enhance distinguishability between nearby states, and improve estimation performance~\cite{rivas2014quantum,breuer2016colloquium,breuer2009measure,de2017dynamics,li2021enhanced,ccakmak2017non}. For this reason, environmental memory is increasingly viewed not only as a modification of noise, but also as a potential resource for quantum metrology~\cite{chin2012quantum,berrada2013non,macieszczak2015zeno,yang2019memory,yang2024control,chen2024quantum,mirkin2020quantum,berrada2026optimizing}.

A complementary route to achieve noise-resilient quantum sensing is engineering the system Hamiltonian, which is capable of displaying robustness against noise \cite{bai2023floquet,o2019hamiltonian,zhou2023robust,zhai2023control}. Specifically, well-designed dynamical evolution can modify the way information about the parameter is imprinted onto the quantum states. In continuous-variable platforms, the inclusion of squeezing-type interactions is especially attractive, as it redistributes quantum fluctuations in phase space and can enhance the sensitivity of the probe without leaving the Gaussian framework. In this work, one of our objectives is to demonstrate that such dynamical evolution can alter the time-scaling of QFI, leading to enhanced precision, especially in the short-time regime where decoherence has not yet fully developed.
Despite substantial progress, the combined impact of squeezing and environmental memory in noisy continuous-variable metrology remains insufficiently understood. In particular, it is unclear when memory effects can complement Hamiltonian engineering, when they compete with it, and which mechanism ultimately provides the larger metrological gain.

In this work, we address these questions through a unified continuous-variable framework for noisy frequency estimation that simultaneously incorporates Hamiltonian engineering and non-Markovian dynamics. In particular,  we incorporate squeezing directly into the system Hamiltonian and demonstrate that it enhances the QFI, including a tunable higher-order correction into the short-time regime, thereby counteracting the Markovian degradation of precision. Using the quantum Brownian motion model, we estimate the frequency under non-Markovian dynamics and show that non-Markovian memory effects can induce a temporary restoration and enhance estimation precision through information backflow. Furthermore, we compare squeezing- and memory-induced gain in estimation, identifying parameter regimes in which each mechanism dominates. We also examine the performance of Gaussian measurement strategies, including homodyne, heterodyne, and optimized general-dyne detection and determine when they approach the ultimate quantum limits, and when more general measurements become advantageous.

This paper is organized as follows. In Sec. \ref{sec:estimationtheoryGaussian}, we introduce the theoretical framework for single parameter estimation in the Gaussian regime, including the formulation of QFI and general-dyne measurement strategies. In Sec.  \ref{sec:squeezingadvantage}, we analyze the role of Hamiltonian engineering by incorporating squeezing into the system Hamiltonian and demonstrate its impact on enhancing the QFI under dissipative dynamics. In Sec. \ref{sec:non-markovianstudy}, we investigate the interplay and competition between non-Markovianity and squeezing within the quantum Brownian motion model, highlighting how memory effects and information backflow can either enhance or limit the estimation precision depending on the parameter regime. In Sec. \ref{sec:cfi}, we examine the achievability of the QFI through Gaussian (dyne) measurements and identify the regimes where they remain optimal. Finally, concluding remarks are presented in Sec. \ref{sec:conclu}.
%, we summarize our main results and outline future research directions.

\section{Single parameter estimation theory in Gaussian framework}
\label{sec:estimationtheoryGaussian}

In quantum estimation theory, the estimation of an unknown parameter begins with the preparation of a probe state \(\rho_{0}\), which is independent of the parameter of interest. This probe is subsequently subjected to a 
%undergoes 
a parameter-dependent dynamical map \(\mathcal{L}_\omega\), resulting in the encoded state \(\rho(\omega)=\mathcal{L}_\omega \rho_{0}\), which carries information about parameter \(\omega\). To extract it, 
%Information about the parameter is then extracted by 
one performs a positive operator-valued measurement (POVM), described by a set of positive semidefinite operators \(\{\Pi_x\}_x\) which satisfies \(\Pi_x \geq 0\) and \(\int \ dx \, \Pi_x = \mathbb{\hat{I}}\), where \(\mathbb{\hat{I}}\) is the identity operator over the underlying Hilbert space. By repeating the measurements \(M\) times, one obtains a statistical sample of independent outcomes \(\mathbf{x}=\{x_{1},\dots ,x_{M}\}\), from which an estimator \(\tilde{\omega}(\mathbf{x})\) is constructed to infer the value of the parameter. The precision of an unbiased estimator is then quantified by its variance \(\mathrm{Var}(\tilde{\omega})\), which is bounded by Cram{\'e}r-Rao inequality \cite{cramer1999mathematical},
\begin{equation}
\mathrm{Var}(\tilde{\omega}) \geq \frac{1}{M F(\omega)}.
\end{equation}
Here, \(F(\omega)\) is the classical Fisher information (CFI), defined as
\begin{equation}
F(\omega) = \int dx \, p(x|\omega) \left[ \partial_\omega \ln p(x|\omega) \right]^2,
\end{equation}
where \(\partial_\omega:=\frac{\partial}{\partial_\omega}\) and \(p(x|\omega) = \mathrm{Tr}\left[\Pi_x \rho(\omega)\right]\) is the conditional probability distribution built from measurement statistics \cite{cramer1999mathematical,rao1945information}. Thus, the CFI determines the precision achievable about \(\omega\) for that specific POVM. 

To determine the ultimate precision allowed by quantum mechanics, one optimizes over all possible POVMs, leading to quantum Fisher information (QFI), \(\mathcal{F}(\omega)\), defined as 
\begin{equation}
\mathcal{F}(\omega) = \max_{\{\Pi_x\}} F(\omega).
\end{equation}
The QFI provides the 
%ultimate bound 
fundamental limit on estimation precision through the quantum Cram{\'e}r-Rao bound
%inequality 
\cite{helstrom1967minimum,helstrom1969quantum,paris2009quantum},
\begin{equation}
\mathrm{Var}(\tilde{\omega}) \geq \frac{1}{M \mathcal{F}(\omega)}.
\end{equation}
An equivalent and widely used %definition 
formulation of QFI is 
%given 
expressed in terms of the symmetric logarithmic derivative (SLD) operator \(L(\omega)\), implicitly via
%defined by
\begin{equation}
2\,\partial_\omega \rho(\omega) = L(\omega)\rho(\omega) + \rho(\omega)L(\omega).
\end{equation}
In terms of the SLD, the QFI reads \cite{braunstein1994statistical,liu2016quantum} 
\begin{equation}
\mathcal{F}(\omega) = \mathrm{Tr}\left[\rho(\omega) L^2(\omega)\right]. 
\end{equation}
The spectral decomposition of the SLD determines the optimal measurement that saturates the quantum Cram{\'e}r-Rao bound i.e., the measurement for which the CFI equals the QFI \cite{braunstein1994statistical,paris2009quantum}. {In the following subsection, we outline the formalism of QFI within the Gaussian framework. A brief overview of Gaussian systems is provided in Appendix~\ref{app:Gaussianframework}.}

\subsection{Quantum Fisher information for Gaussian states }

For Gaussian states, the QFI, \(\mathcal{F}(\omega)\), admits a compact analytical expression entirely in terms of first and second moments, $\avg{\hat{\bX}}$ and $\sigma$ \cite{vsafranek2015quantum,vsafranek2019estimation}:
\begin{eqnarray}
 \mathcal{F}(\omega) &=& 2\left(\frac{\partial \avg{\hat{\bX}}}{\partial\omega}\right)^\top \bsigma^{-1} \left(\frac{\partial \avg{\hat{\bX}}}{\partial\omega}\right) \nonumber\\
&+& \frac{1}{2}\text{vec}\left[\frac{\partial\bsigma}{\partial\omega}\right]^\top \mathcal{M}^{-1} \text{vec}\left[\frac{\partial\bsigma}{\partial\omega}\right],
\label{eq:qfi_general}   
\end{eqnarray}
where \(\mathcal{M} = \bsigma \otimes \bsigma - \Omega \otimes \Omega \), and \(\text{vec}[.]\) denotes vectorization of a matrix, with \(\Omega\) representing the symplectic form. Note that this expression separates  the variations in the displacement vector from the changes in the covariance matrix. 

We now turn to a broad class of Gaussian measurements known as general-dyne detections. These measurements extend the notion of phase-space measurements and can be understood as a generalized form of over-completeness relation satisfied by coherent states \cite{ferraro2005gaussian,genoni2016conditional,Serafini_2017}. A general-dyne measurement is defined through a Gaussian POVM, constructed from a reference Gaussian state \(\hat{\rho}_{m}\) with vanishing first moments and covariance matrix \(\bsigma_{m}\). The completeness relation reads
\begin{equation}
    \frac{1}{(2\pi)^d} \int d^{2d}\mathbf{r}_m \,
\hat{D}^\dagger_{\mathbf{r}_m}\,\hat{\rho}_m\,\hat{D}_{\mathbf{r}_m} = \hat{I},
\end{equation}
where \(\hat{D}_{\mathbf{r}_m}\) is the displacement operator and \(\mathbf{r}_m\) labels the measurement outcomes. The corresponding POVM elements are given by
\begin{equation}
    \hat{\Pi}_{\mathbf{r}_m} = \frac{\hat{D}^\dagger_{\mathbf{r}_m}\,\hat{\rho}_m\,\hat{D}_{\mathbf{r}_m}}{(2\pi)^d}.
\end{equation}
The measurement is completely specified by the covariance matrix \(\bsigma_{m}\) of the seed state. When \(\hat{\rho}_{m}\) is pure, the measurement is said to be ideal and this formalism naturally includes the standard detection schemes. For example, in the single-mode case, homodyne detection of the quadrature \(\hat{q}\) is obtained in the limit of infinite squeezing,
\begin{equation}
\bsigma_m = \lim_{s \to -\infty}  \begin{pmatrix}
e^{2s} & 0 \\
0 & e^{-2s}
\end{pmatrix}=\lim_{z \to 0}  \begin{pmatrix}
z & 0 \\
0 & z^{-1}
\end{pmatrix},
\end{equation}
while heterodyne detection corresponds to \(\bsigma_{m}=\mathbb{I}_2\). When a general-dyne is performed on a Gaussian state \(\hat{\rho}\) with displacement vector \(\langle \hat{\mathbf{X}} \rangle\) and covariance matrix \(\bsigma\), the resulting outcome statistics remains  Gaussian with covariance matrix \(\bSigma = (\bsigma+\bsigma_{m})/2\). The sensitivity of such measurements to the parameter \(\omega\) is quantified by the CFI which for Gaussian distributions, 
%one has the expression of \(F(\omega)\) 
takes the form as \cite{monras2013phase}
\begin{eqnarray}
 F(\omega) &=& \left(\frac{\partial \avg{\hat{\bX}}}{\partial\omega}\right)^\top \bSigma^{-1} \left(\frac{\partial \avg{\hat{\bX}}}{\partial\omega}\right) \nonumber\\
&+& \frac{1}{2} \mathrm{Tr}
\left[\bSigma^{-1} \left(\frac{\partial \bSigma}{\partial \omega}\right) \bSigma^{-1} \left(\frac{\partial \bSigma}{\partial \omega}\right)\right].
\label{eq:cfi_general}   
\end{eqnarray}
We will employ the above expression to demonstrate the advantage of squeezing in enhancing parameter estimation precision.

\section{Advantage of introducing squeezing in the Hamiltonian}
\label{sec:squeezingadvantage}

Let us consider a single-mode harmonic oscillator with Hamiltonian
\begin{equation}
    H=\omega a^\dagger a,
    \label{eq:Ham_withoutsqueezing}
\end{equation}
coupled to a thermal bath. 
Our goal is to estimate the frequency parameter $\omega$, which is encoded in a probe system initially prepared in the coherent state $\ket{\alpha}$ with $\alpha=|\alpha|e^{i\theta}$. The encoding is performed through the channel parameterization governed by the 
%dynamical 
Gorini Kossakowski Sudarshan Lindblad master equation \cite{breuer2002theory}
\begin{eqnarray}
    \frac{d\rho}{dt}
= -i[H, \rho]
+ \frac{\gamma}{2}(n+1)\,\mathcal{L}[a](\rho)
+ \frac{\gamma}{2}n\,\mathcal{L}[a^\dagger](\rho),
\label{eq:evolution}
\end{eqnarray}
where the Lindblad operator is given by
\begin{eqnarray}
    \mathcal{L}[A](\rho)
&=& A \rho A^\dagger
- \frac{1}{2} A^\dagger A \rho
- \frac{1}{2} \rho A^\dagger A,
\end{eqnarray}
with $\gamma$ denoting the damping rate and $n$ representing the mean thermal photon number of the bath.

\begin{figure*}
\includegraphics[width=1.0\linewidth]{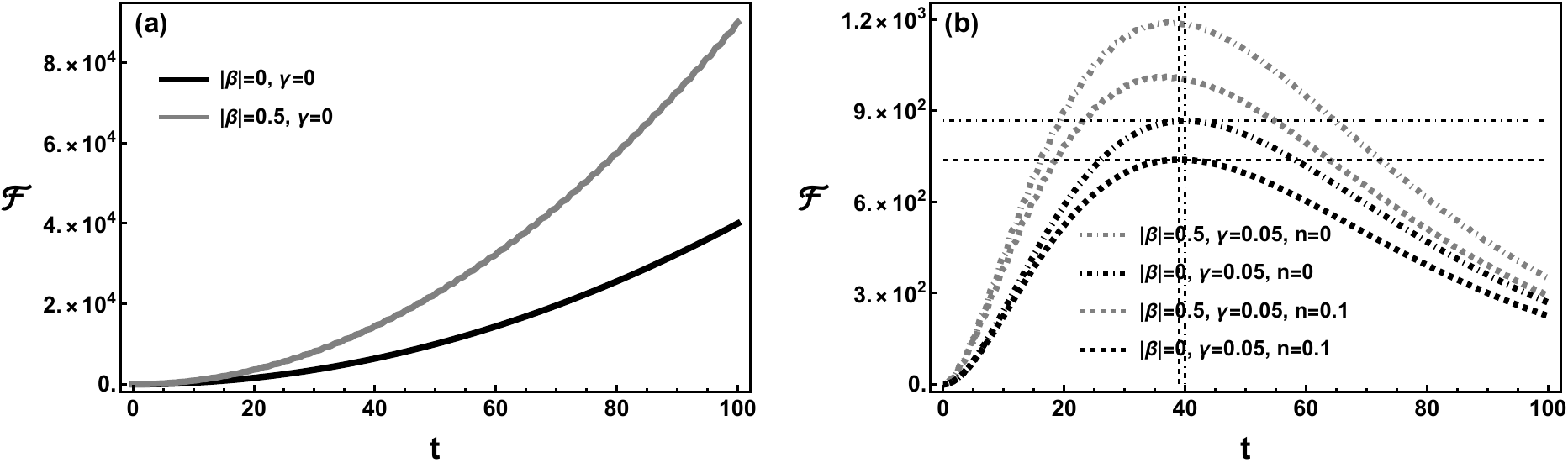}
    \caption{QFI as a function of time. In the left panel (a), $\gamma = 0$; the black (continuous) and gray (continuous) curves correspond to $|\beta| = 0$ and $|\beta| = 0.5$, respectively. In the right panel (b), $\gamma = 0.05$. Other parameters are fixed at $\omega = 2.1$, $\beta = -0.5i$, and $\alpha = i$.}
\label{fig:nonzerobeta_nonzerogamma}
\end{figure*}

The quantum Fisher information as a function of time $t$ can then be expressed in the compact form as (see Appendix \ref{app:QFI_computation})
\begin{equation}
    \mathcal{F}(\omega,t)=\frac{4 |\alpha| ^2 t^2}{2 n \left(e^{\gamma  t}-1\right)+e^{\gamma  t}}.
    \label{eq:QFI_betazero}
\end{equation}
Note that in the case of unitary evolution, governed solely by the Schr\"odinger equation with the Hamiltonian $H$, the QFI simplifies to $\mathcal{F}(\omega,t)=4|\alpha|^2 t^2$, which is a monotonically increasing quadratic function of time $t$ (see Fig.~\ref{fig:nonzerobeta_nonzerogamma}(a)). Upon introducing a vacuum bath (i.e., $n=0$, \(\gamma \ne 0\)) with damping rate $\gamma$, the QFI takes the form as $\mathcal{F}(\omega,t)=4|\alpha|^2 t^2 e^{-\gamma t}$, which indicates that, although the QFI initially grows with time, it eventually undergoes exponential decay beyond a certain time due to the presence of the environment.  In particular, the QFI attains its maximum at $t = 2/\gamma$, after which it decays exponentially. Furthermore, in the presence of a thermal bath with a non-vanishing photon number (\(n>0\)), the qualitative behavior of the QFI remains largely similar. Specifically, the QFI decreases with increasing $n$, while still exhibiting an initial growth followed by an exponential decay after a certain time, as shown in  Fig.~\ref{fig:nonzerobeta_nonzerogamma}(b). 
%In Fig.~\ref{fig:nonzerobeta_nonzerogamma}, we present the time evolution of the QFI. In the left panel, the lower curve (blue) corresponds to $\beta = 0$. In the right panel, the two lower curves represent $\beta = 0$, with $N = 0.1$ (red) and $N = 0$ (orange). For both panels, we fix $\alpha = i$, while in the right panel we set $\gamma = 0.05$. In the right panel, the optimal time corresponding to $\beta = 0$ with $N = 0$ is found to be $t = 40$ \ayan{what is the unit?}, and is slightly smaller for $N = 0.1$. In both cases, the QFI exhibits a similar behavior -- it initially increases, reaches a maximum, and then decays exponentially beyond a certain time.

{\it Beneficial effect of squeezing on parameter estimation. } We now examine the impact of incorporating a nonlinear optical framework into the system to explore whether it provides any advantage in achieving a higher QFI, especially in the presence of decoherence. In this context, a wide range of such approaches has been explored in the literature, including higher-order terms in the field operators within the Hamiltonian~\cite{guo_arXiv_2025}, as well as Kerr-type nonlinearities~\cite{Genoni_PRA_2009,manju2026quadratic,Rossi_PRA_2016,Notarnicola_APLQ_2024,Xiao_QIP_2020,Chang_PRA_2022}. In the present work, we instead focus on incorporating squeezing into the governing Hamiltonian, which is given by
\begin{equation}
H=\omega a^\dagger a+\beta a^{\dagger 2}+\beta^{*}a^2,
\label{eq:Ham_withsqueezing}
\end{equation}
where $\beta=|\beta|e^{i\delta}$ denotes the squeezing parameter of the system. Note importantly that the inclusion of quadratic terms in the Hamiltonian preserves the Gaussian nature of the framework. Consequently, we can effectively exploit the Gaussian phase-space formalism for our analysis. In contrast, introducing higher-order field operators would necessarily compel us to work within a non-Gaussian framework (cf. \cite{Adesso_2014,Serafini_2017}).

In contrast to the case of vanishing $\beta$, a compact analytical expression for the QFI is not available in general. However, under certain approximations, analytical expressions can still be obtained. For instance, in the short-time regime ($t \ll 1$), the QFI admits the analytical form as
\begin{eqnarray}
   \mathcal{F}(\omega,t)= 4 |\alpha| ^2 t^2 \Big(1-\gamma t(1+2n)+4 |\beta|  t \sin (\delta -2 \theta )\Big).\nonumber\\
   \label{eq:QFI_betanonzero}
\end{eqnarray}
It is therefore evident that the squeezing phase $\delta$ can be tuned such that the last term in Eq.~\eqref{eq:QFI_betanonzero} remains positive, thereby yielding a cubic enhancement with respect to time over the existing quadratic scaling of the QFI. %Moreover,
Further analytical progress can be made in the regime of low squeezing $\beta$ ($|\beta| \ll 1$)
%, we obtain a compact expression for the QFI 
under unitary evolution (i.e., $\gamma = 0$). Expanding the QFI up to second order in $\beta$, 
the QFI is given by
\begin{align} 
\nonumber \mathcal{F}(\omega,t)= &  4 |\alpha| ^2 t^2 + 16 t |\alpha|^2\frac{|\beta |}{\,\,\omega^2} 
\Big\{\omega t \cos (\delta -2 \theta )\\
& \qquad \qquad -\sin ( \omega t)
\cos (\delta -2 \theta +\omega t)\Big\}\nonumber \\
&+
4\, \frac{|\beta|^2}{\,\omega^4} \Bigg\{1+2|\alpha|^2+2t^2\omega^2 \big(1+5|\alpha|^2\big) \nonumber \\ 
&\quad -\cos(2t\omega)\Big[1+|\alpha|^2\big(2+6t^2\omega^2\big)\Big] \nonumber \\ 
&\quad -2t\omega\sin(2t\omega)\big(1+2|\alpha|^2\big)\Bigg\}\,.\label{eq:QFI_betanonzero2}
\end{align}
Here again, one can observe that the 
%can be tuned such that
squeezing phase $\delta$ ensures that the coefficients of both $|\beta|$ and $|\beta|^2$ in Eq.~\eqref{eq:QFI_betanonzero2} can be made positive, leading to an enhancement of the QFI compared to the case $\beta = 0$. 

In contrast, 
%for non-zero $\gamma$,
for dissipative dynamics (\(\gamma \ne 0\)), no such compact analytical expression 
%exists 
is available even for the small $\beta$ approximation; therefore, the QFI must be evaluated numerically. To elucidate the beneficial role of squeezing, we compare the time evolution of the QFI in the presence of a bath for both vanishing and nonvanishing values of the system squeezing parameter, as depicted in Fig.~\ref{fig:nonzerobeta_nonzerogamma}(b). The comparison clearly demonstrates that squeezing effectively mitigates the detrimental effects induced by the environment. Specifically,  a non-zero $\beta$ consistently yields a higher QFI throughout the evolution, although the overall qualitative behavior of QFI in the presence of squeezing remains similar to the $\beta = 0$ case over time. This indicates that 
%the presence of 
squeezing can partially counteract the damping effects induced by environmental interaction. However, 
beyond a certain time, dissipative processes dominate, and the QFI inevitably undergoes exponential decay, regardless of the strength of the nonlinearity introduced into the system.

%\ayan{Should we give small-$\beta$ approximated QFI plots for nonzero $\gamma$? Since we don't have the compact expression of the small-$\beta$ approximated QFI for non-vanishing $\gamma$, we need to employ numerical techniques. But, on the other hand, we already provide QFI plots with respect to non-zero $\beta$ and $\gamma$. Our aim was to provide some compact analytical expressions for QFI under certain assumptions, but if it is not possible, then why one would give any plot made from numerically obtained data set that too with approximation where already without approximated QFI plots are available.}

%In Fig.~\ref{fig:nonzerobeta_nonzerogamma}, we plot the QFI as a function of time. The left panel illustrates its evolution under purely unitary dynamics governed by the Hamiltonian in Eq.~\eqref{eq:Ham_withsqueezing}, for $\beta = 0$ (blue) and $\beta = 0.5$ (orange). The right panel shows the corresponding behavior in the presence of a thermal bath, with dynamics governed by Eq.~\eqref{eq:evolution} using the same Hamiltonian. In this case, the lower two curves correspond to $\beta = 0$ with $N = 0$ (orange) and $N = 0.1$ (red), while the upper two correspond to $\beta = 0.5$ with $N = 0$ (blue) and $N = 0.1$ (green). For both panels, we fix $\alpha = i$ and $\omega = 2.1$, and in the right panel we set $\gamma = 0.05$.

\section{Competition between non-Markovianity and squeezing in parameter estimation
}
\label{sec:non-markovianstudy}

In the previous section, we analyzed frequency estimation under Markovian noise, which assumes a memoryless environment, and showed that suitable Hamiltonian engineering, particularly via squeezing, can substantially enhance precision even in the presence of decoherence.
%In the previous section, we studied frequency estimation under Markovian dynamics and showed how suitable Hamiltonian engineering can improve the estimation precision. The Markovian description assumes 
%that the environment has no memory, 
%a memoryless environment, resulting in a continuous and irreversible loss of quantum resources. 
In  realistic settings, however, reservoirs often exhibit non-negligible memory effects, 
%structured environments, 
even in the weak-coupling regimes~\cite{vasile2009continuous} or at high temperatures~\cite{torre2018exact}. %exhibit memory effects that 
Such situations require a non-Markovian description, where the system dynamics is influenced by its dynamical history. A key feature of non-Markovianity is the possibility of temporary backflow of information from the environment to the system, which can partially restore quantum coherence and enhance the distinguishability of quantum states. This suggests that non-Markovianity can be viewed as a useful resource for quantum estimation,  \cite{porto2025temperature,gaidi2025impact}, rather than merely a source of noise.

Motivated by these observations, we now investigate the interplay between non-Markovian memory effects and squeezing-induced Hamiltonian engineering in frequency estimation. In particular, we ask whether the environmental memory can complement or even surpass the enhancement provided by squeezing in the presence of noise. To address this question, we employ the paradigmatic model of quantum Brownian motion (QBM), describing a quantum harmonic oscillator coupled to a bosonic thermal reservoir. This model naturally captures the combined effects of dissipation, diffusion, and memory effects, thereby providing an ideal set-up to compare noise-assisted benefit with control-based strategies.  
%and to assess how these mechanism compete with, or reinforce, the metrological advantage arising from squeezing in frequency estimation. 
The total Hamiltonian describing the system,  environment and system-bath interaction is given by \cite{intravaia2003density}
\begin{equation}
H = \frac{\omega^2}{2 } (\hat{p}^2+\hat{q}^2)
+ \sum_n \left( \frac{\hat{P}_n^2}{2m_n} + \frac{1}{2} m_n \omega_n^2 \hat{Q}_n^2 \right)
+ \xi \hat{q} \sum_n K_n \hat{Q}_n,
\end{equation}
where $(\hat{q}, \hat{p})$ denotes the canonical position and momentum operators of the system oscillator with frequency $\omega$, while $(\hat{Q}_n, \hat{P}_n)$-pair describes the environmental modes with frequency \(\omega_n\). The coupling constants $K_n$ and the global parameter $\xi$ determine the strength of the system-bath interaction.

The environment is fully characterized by the spectral density function
\begin{equation}
J(\Lambda) = \sum_n \frac{K_n^2}{2m_n \omega_n} \delta(\Lambda - \omega_n),
\end{equation}
which we consider to be of Ohmic type with Lorentz--Drude cutoff \cite{breuer2002theory,maniscalco2004lindblad},
\begin{equation}
J(\Lambda) = \frac{2\Lambda}{\pi} \frac{\Lambda_c^2}{\Lambda_c^2 + \Lambda^2},
\label{Ohmic}
\end{equation}
with $\Lambda_c$ being the cutoff frequency defining the bath correlation time $\tau_B \sim \Lambda_c^{-1}$.

Under the assumptions of weak coupling ($\xi \ll 1$), initially factorized states, and the secular approximation, the reduced dynamics of the system is governed by a time-local Gorini Kossakowski Sudarshan Lindblad master equation of the form \cite{intravaia2003density,intravaia2003comparison,frigerio2021exploiting}
\begin{align}
\frac{d\rho}{dt} &= \frac{\Delta(t) + \gamma(t)}{2} 
\left( 2 \hat{a} \rho \hat{a}^\dagger - \hat{a}^\dagger \hat{a} \rho - \rho \hat{a}^\dagger \hat{a} \right) \nonumber \\
&\quad + \frac{\Delta(t) - \gamma(t)}{2} 
\left( 2 \hat{a}^\dagger \rho \hat{a} - \hat{a} \hat{a}^\dagger \rho - \rho \hat{a} \hat{a}^\dagger \right).
\label{master_equ}
\end{align}
The time-dependent coefficients $\gamma(t)$ and $\Delta(t)$ describe dissipation and diffusion, respectively, and they fully encode 
the system-environment interaction. In the case of a thermal environment at temperature $T$, the coefficients $\Delta(t)$ and $\gamma(t)$ read respectively as
%take the form 
\cite{intravaia2003comparison}
\begin{equation}
\Delta(t) = \xi^2 \int_0^t ds \int_0^\infty d\Lambda \, J(\Lambda)\, \coth\!\left(\frac{\hbar \Lambda}{2 k_{B} T}\right) \cos(\Lambda s)\cos(\omega s),
\end{equation}
and 
\begin{equation}
\gamma(t) = \xi^2 \int_0^t ds \int_0^\infty d\Lambda \, J(\Lambda)\, \sin(\Lambda s)\sin(\omega s).
\end{equation}
Note that \(\coth\!\left(\frac{\hbar \Lambda}{2 k_{B} T}\right) = 2 n(\Lambda) + 1\),
where $n(\Lambda)$ is the mean number of thermal photons with frequency $\Lambda$, and $\omega$ is the characteristic frequency of the system. For the chosen spectral density, the dissipation coefficient admits the analytic expression \cite{maniscalco2004lindblad}
\begin{equation}
\gamma(t) = \frac{\xi^2 \omega r^2}{2(r^2 + 1)} 
\left[ 1 - e^{-\Lambda_c t} \cos(\omega t) - r e^{-\Lambda_c t} \sin(\omega t) \right],
\end{equation}
where $r = \Lambda_c / \omega$ which represent the degree of non-Markovianity.

In the high-temperature regime ($k_{B}T \gg \hbar \Lambda_{c}, \hbar \omega$), the diffusion coefficient reads \cite{maniscalco2004lindblad,caldeira1983path}
\begin{equation}
\Delta(t) = \frac{2 \xi^2 k_{B}T r^2}{r^2 + 1} 
\left[ 1 - e^{-\Lambda_c t} \left( \cos(\omega t) - \frac{1}{r} \sin(\omega t) \right) \right].
\end{equation}

In general, after a time ($t \approx \Lambda_c^{-1}$), the coefficients approach their Markovian stationary values, and the system behaves according to the predictions of the Markovian approximation \cite{vasile2009continuous}
\begin{equation}
\gamma_{M} = \frac{\xi^2 \omega r^2}{(r^2 + 1)}, 
\qquad
\Delta_{M} = \frac{2 \xi^2 k_{B}T r^2}{r^2 + 1}.
\end{equation}

The dynamical map associated with the above master equation is completely positive for all times. However, it is \emph{CP-divisible} if the time-dependent rates satisfy
\begin{equation}
\Gamma_\pm(t) = \Delta(t) \pm \gamma(t) \geq 0 \quad \forall t.
\end{equation}

\begin{figure*}
\includegraphics[width=1.0\linewidth]{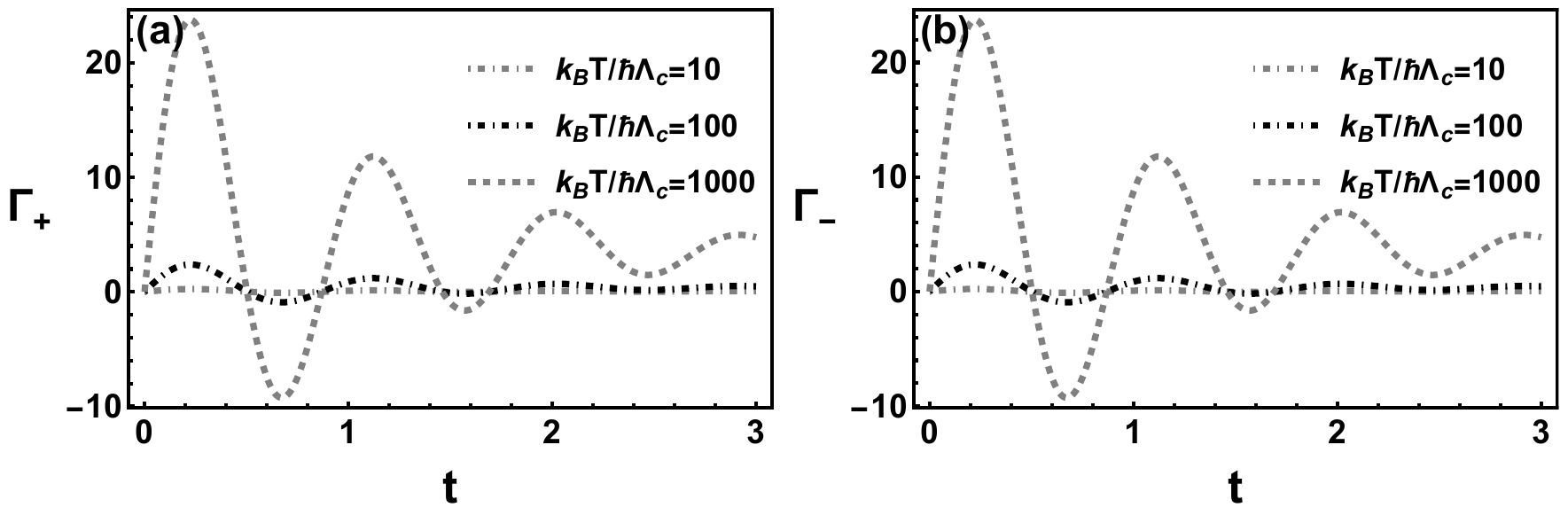}
    \caption{(a) $\Gamma_+(t) = \Delta(t) + \gamma(t)$ (ordinate) and (b) $\Gamma_+(t) = \Delta(t) - \gamma(t)$  (ordinate)  as a function of time (abscissa). The other parameter values are $\omega = 7$, $\xi = 0.3$, and $\Lambda_{c} = 1$.}
    \label{fig:diffusion}
\end{figure*}
Temporary negativity of $\Gamma_\pm(t) = \Delta(t) \pm \gamma(t)$ signals a breakdown of CP-divisibility and hence the emergence of non-Markovian dynamics \cite{breuer2009measure,laine2010measure,torre2015non}. Physically, this corresponds to a backflow of information from the environment to the system, induced by the finite memory time of the bath. Moreover, for the specific spectral density, we consider in Eq. (\ref{Ohmic}), the non-Markovianity of the QBM channel is governed by the ratio $r$ \cite{maniscalco2004lindblad,vasile2011quantifying,torre2015non}. In particular, $r \gg 1$: Markovian regime (fast bath, negligible memory), $r \sim 1$: intermediate regime having memory effect and $r \ll 1$: strongly non-Markovian regime with pronounced oscillations. Since the QBM channel preserves Gaussianity, the system state remains Gaussian throughout the evolution. It is therefore fully characterized by the first moments and the covariance matrix.

\begin{figure*}
\includegraphics[width=1.0\linewidth]{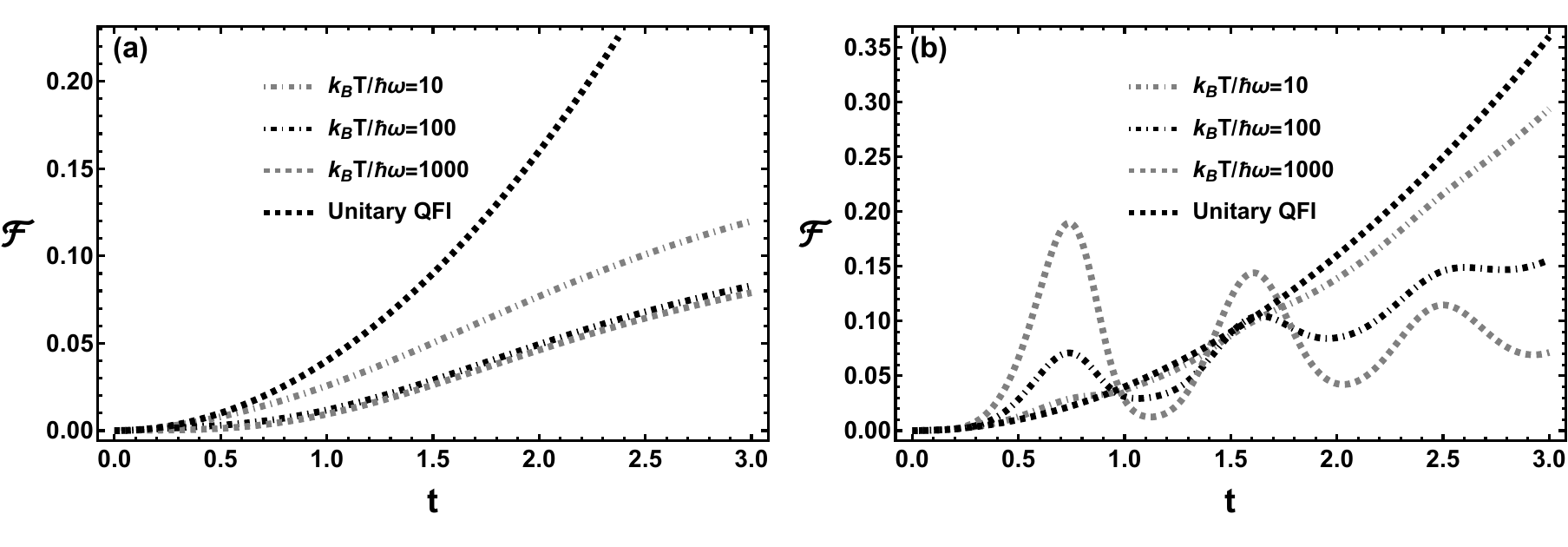}
    \caption{ QFI (ordinate) against time (abscissa) under unitary, non-Markovian and Markovian dynamics without squeezing ($\beta =0$). The left panel (a) corresponds to  $\Lambda_{c}=2$ and $\omega=1$ (Markovian evolution). For right panel (b),  $\Lambda_{c}=1$ and $\omega=7$ with $\xi = 0.3$ and $\alpha=0.1$ (Non-Markovian dynamics). }
    \label{fig:Non_Markovian}
\end{figure*}

\begin{figure}
\includegraphics[width=1.0\linewidth]{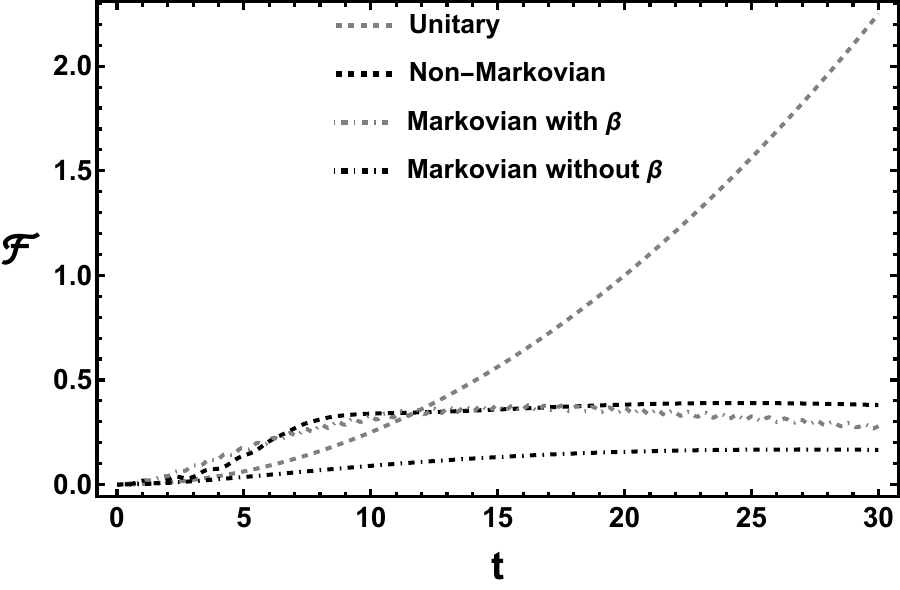}
    \caption{Competition between Hamiltonian engineering and non-Markovianity in parameter estimation.  QFI (ordinate) with respect to time (abscissa) when squeezing is introduced, i.e.,  \(\beta \neq 0\). The parameters are $\xi = 0.3$, $\Lambda_{c}=1$, $\omega=5$, $\beta=0.2$, $\alpha=0.025$, $k_{B}T/\hbar \Lambda_{c}=4$, $\gamma = 0.07$ and $n=0.4$. There exists a time region in which \(\mathcal{F}\) with \(\beta  \neq 0\) under Markovian dynamics can be higher than that of non-Markovian evoluion without \(\beta\). It shows that the Hamiltonian engineering can  sometime be more beneficial than the non-Markovian dynamics.  }
    \label{fig:Non-Markovian-Squeezing}
\end{figure}
For an initial coherent state \(\sigma_{0}= \mathbb{I}\), the mean vector and the covariance matrix evolves as \cite{vasile2009continuous,hesabi2023memory}
\begin{equation}
\langle \hat{X}(t)\rangle = S(t) X(0),~~~
\sigma(t) = e^{-\Gamma(t)} \sigma(0) + \mathcal{T}(t) \mathbb{I}\,
\end{equation}
where \(S(t)\) and \(\mathcal{T}(t)\) are \(2 N\times 2N\), with $N$ being the total number of system modes, are real matrices that characterize the exact evolution of Eq. (\ref{master_equ}), given by \cite{maniscalco2004lindblad}
\begin{equation}
S(t) = e^{-[\Gamma(t)]} R(t), \quad   \mathcal{T}(t) = e^{-\Gamma(t)} \Delta_{\Gamma}(t),
\end{equation}
with \(\Gamma(t) = 2\int_0^t \gamma(s)\, ds, \quad \Delta_{\Gamma}(t)= \int_0^t e^{\Gamma(s)} \Delta(s) ds \)
and $R(t)$ is the phase-space rotation matrix by angle \(\omega t \). We are especially interested in the short-time dynamics that are non-Markovian, then in this regime and within weak-coupling limit, one can expand the exponential terms appearing in \(\mathcal{T}(t)\) in Taylor series. In this case, one can approximate \(\mathcal{T}(t)\) as \cite{vasile2009continuous}\begin{equation}
\mathcal{T}(t) \simeq \int_0^t \Delta(s)\, ds,
\end{equation}
which is valid in the weak-coupling limit. Now, using Eq. (\ref{eq:qfi_general}), we obtain the corresponding QFI for non-Markovian dynamics of QBM to study the beneficial role of non-Markovianity for frequency estimation. To access the impact of memory effects on frequency estimation, we compare the QFI under Markovian \((r\gg1)\) and non-Markovian \((r\ll1)\) dynamics while keeping all system parameters fixed. 
%The degree of non-Markovianity is governed by the ratio \(r\), which sets the bath correlation time. 
In the Markovian limit, the environment has negligible memory and induces a monotonic loss of information, whereas for \(r\ll1\), finite correlation time gives rise to memory effects. This distinction is clearly reflected in the evolution of time dependent coefficients $\Gamma_\pm(t)$ shown in Fig.~\ref{fig:diffusion}, which exhibit transient oscillations and temporal negativity for \(r\ll1\) ensuring the breakdown of CP-divisibility and presence of memory effects.\\

The metrological consequences of these effects are illustrated in Fig.~\ref{fig:Non_Markovian}.
%The two panels illustrate the behavior of the QFI for different temperature values in distinct dynamical regimes characterized by the parameter \(r\). The left panel corresponds to the \(r=2\), in this case,
In the Markovian domain, as expected, the QFI exhibits a smooth, monotonic behavior -- it initially grows and then gradually deviates from the unitary curve due to environmental noise. As the temperature increases, the QFI is progressively suppressed, reflecting the dominant role of thermal decoherence. The absence of oscillations or non-monotonic features indicates that information continuously flows from the system to the environment without any backflow. In contrast, 
%the right panel corresponds to the \(r=0.15\), where 
in the non-Markovian domain (as shown in the right panel of Fig. \ref{fig:Non_Markovian}), 
%the environment possesses a long memory time. Here,
the QFI displays noticeable deviations from the Markovian behavior, including mild non-monotonic features and for certain time intervals, the QFI exceeds the unitary curve, which can be a clear signature of information backflow from the environment to the system. This enhancement is directly linked to the temporal structure of the diffusion and dissipation coefficients, whose transient behavior allows partial recovery of quantum coherence. Furthermore, the effect of temperature remains significant, but its impact is mitigated by memory effects, leading to a less pronounced suppression of QFI compared to the Markovian case. Overall, the comparison demonstrates that while Markovian dynamics lead to irreversible loss of information, non-Markovianity modifies the decoherence process by introducing memory-induced corrections. These effects show the temporarily enhanced estimation precision, highlighting the role of environmental memory as a useful resource in quantum metrology. \\

In contrast to the unitary scenario, where the parameter \(\omega\) is encoded solely through phase rotation generated by the Hamiltonian (\ref{eq:Ham_withoutsqueezing}),  non-Markovian dynamics considered here modify the encoding mechanism itself. Specifically, the evolved state is no longer of the form \(\rho(\omega,t)=U \rho_{0}U^{\dagger}\), but instead arises from a parameter-dependent dynamical map \(\rho(\omega,t) = \mathcal{E}_\omega^{\,t}[\rho_0]\), where the map implicitly depends on \(\omega\) through the time-dependent coefficients \(\gamma (t)\) and \(\Delta (t)\). As a result, the \(\omega\)-dependence is distributed across both the first moments and the covariance matrix, leading to a richer encoding structure compared to the purely unitary case. This modified encoding mechanism allows, in certain time intervals, the QFI under non-Markovian dynamics to exceed the corresponding to the unitary value. This extended encoding can generate temporary metrological advantages for certain time intervals. Importantly, this enhancement does not indicate a violation of fundamental precision limits, but rather reflects the fact that the environment, through its finite memory, effectively participates in the  encoding process. The temporary information backflow  manifest as an additional \(\omega\)-sensitive feature in the state evolution, thereby providing an effective resource that enhances estimation precision beyond what is achievable through purely unitary dynamics at the same time. In this context, it is important to note that the observed non-Markovian enhancement depends sensitively on the probe parameters and the observed advantage is more pronounced for 
%small values of 
weak coherent amplitude \(\alpha\) for small time, where the relative contribution of environmental backflow becomes comparable to, and can even dominate over, the intrinsic unitary encoding. In contrast, for large \(\alpha\), the unitary contribution dominates the encoding process, thereby reducing the relative benefit of memory effects.

Let us now concentrate on ths comparison between squeezing-assited Markovian estimation and non-Markovian one (see Fig.~\ref{fig:Non-Markovian-Squeezing}). It reveals that non-Markovianity can outperform squeezing only in a restricted regime of small squeezing strength \(\beta\) and small coherent amplitude \(\alpha\).
%, non-Markovian dynamics can outperform the squeezing-assisted Markovian strategy, indicating that the environmental memory provides an advantage than that Hamiltonian engineering in this parameter regime. 
Nevertheless, as either \(\beta\) or \(\alpha\) increases, the advantage generated by Hamiltonian engineering rapidly surpasses the memory-induced gain. This demonstrates that while non-Markovianity can temporarily mitigate decoherence, squeezing offers a stronger, more robust, and experimentally controllable route to enhanced precision, even in purely Markovian environments.

%the squeezing-induced enhancement becomes more dominant, and the relative advantage arising from non-Markovianity is progressively diminished. These observation suggest that, while non-Markovian memory can provide a transient advantage in specific parameter regime, Hamiltonian engineering through squeezing offers a more robust and controllable stratergy for mitigating detrimental effects of noise in frequency estimation.}\\

%Fig.~\ref{fig:dyne_Non_Markovian} illustrate CFI corresponding to different Gaussian measurement strategies, namely homodyne, heterodyne, and optimized general-dyne detection, alongside the QFI under non-Markovian dynamics. 

Importantly, if one extends the non-Markovian analysis by incorporating squeezing directly in the system Hamiltonian (as discussed in the Sec. \ref{sec:squeezingadvantage}), the dynamics generates the non-zero position momentum correlations, i.e., non-zero off-diagonal terms in the covariance matrix. In such scenario, the phase-space distributions become anisotropic and rotated, implying that the information about the parameter is encoded in correlated quadrature. Under Markovian dynamics, these correlations are rapidly washed out due to irreversible decoherence. In contrast, Non-Markovian dynamics can temporarily preserve and even revive these correlations through information backflow, thereby making the metrological advantage much more pronounced \cite{porto2025temperature}.

\section{Are dyne measurements enough? -- Achievability through dyne measurements }
\label{sec:cfi}

%In this section, we investigate the achievability of the QFI via general dyne measurements. In particular, 
We now examine whether, and under what conditions, the CFI obtained from dyne measurements can saturate the QFI bound. Let us first analyze the CFI in the absence of squeezing in the Hamiltonian, i.e., for \(\beta=0\). As discussed in Sec. \ref{sec:estimationtheoryGaussian}, a general dyne measurement, parametrized by the real parameter $z$, provides a unified framework encompassing homodyne and heterodyne measurement. 
%can be exploited to estimate the encoded parameter. 
Using Eq. \eqref{eq:cfi_general}, we obtain the corresponding expression for the CFI as
\begin{eqnarray}
  F(\omega,t,z)&=&4 t^2 |\alpha|^2\Big(\frac{z \cos^2(wt-\theta)}{(2 n z+z+1) e^{\gamma  t}-2 n z}+\nonumber\\&&\frac{\sin^2(wt-\theta)}{(2 n+z+1) e^{\gamma  t}-2 n}\Big).
\end{eqnarray}
This expression explicitly shows how the estimation precision depends on both the measurement parameter \(z\) and the phase-space orientation of the initial coherent state parameters \(\alpha = |\alpha|e^{i\theta}\). 

 The CFI for homodyne measurement is recovered in the limits \(z \to 0\) for the position-quadrature and \(z \to \infty\) for the momentum-quadrature, while the CFI for heterodyne measurement is obtained for \(z=1\). By optimizing over \(z \in [0,\infty]\), we find that the optimal dyne measurement can, in certain cases, attain the QFI. For example, at \(t=\frac{2\theta + m\pi}{2\omega}\) (\(m=0,1,2, \ldots\)), the CFI reaches the QFI value for even \(m\) under momentum-quadrature measurement \((z\to\infty)\), and for odd \(m\) under position-quadrature measurement \((z \to 0)\), where \(\theta\) denotes the argument of \(\alpha\). This demonstrates that, in the absence of squeezing, Gaussian measurement can achieve the ultimate quantum limit at appropriately chosen times. 

\begin{figure}
    \centering
    \includegraphics[width=1.0\linewidth]{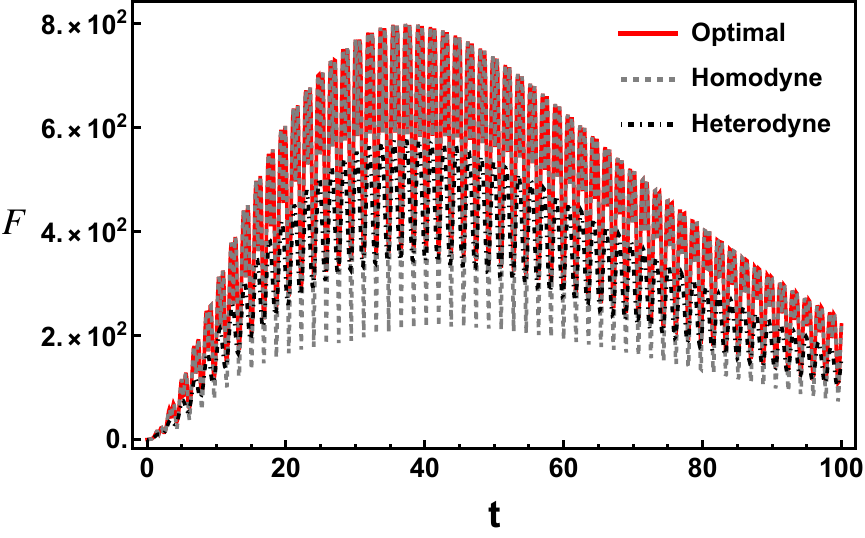}
    \caption{Classical Fisher information (CFI) as a function of time. The CFI corresponding to homodyne and heterodyne measurements are shown in gray (dashed) and black (dot-dashed), respectively, while the dyne-optimal CFI is depicted by the red (continuous) curve. The dyne-optimal CFI coincides with the maximum of the homodyne and heterodyne curves. The parameters are $\omega = 2.1$, $\gamma = 0.05$, $n = 0.1$, and $\beta = 0.3 - 0.5i$, with $\alpha = i$.}
    \label{fig:optdyne}
\end{figure}

Let us now turn to the case of non-vanishing squeezing, $\beta \ne 0$. In contrast to the previous scenario, the analytical evaluation of the CFI with \(\beta \ne 0\) is more involved and consequently,  a compact closed-form expression 
is not available.
%in this regime. 
Nevertheless,  our numerical %analysis 
optimization over the measurement parameters indicates that the optimal dyne measurement corresponds either to a homodyne or a heterodyne measurement, as illustrated in Fig.~\ref{fig:optdyne}. In most instances, homodyne measurement remains optimal, while heterodyne detection seldom becomes optimal. Similar to the $\beta = 0$ case, the dyne-optimized CFI exhibits a pulsating behavior in time, with an envelope that is non-monotonic -- it initially increases and subsequently decays exponentially as a consequence of dissipation governed by the damping rate $\gamma$.
%due to the damping factor 
%Note that although Fig.~\ref{fig:optdyne} is presented for the parameter values $\omega = 2.1$, $\gamma = 0.05$, $n = 0.1$, and $\beta = 0.3-0.5i$ with $\alpha = i$, we observe that similar inferences hold across other parameter values within the feasible range. 

\begin{figure}
    \centering
    \includegraphics[width=1.0\linewidth]{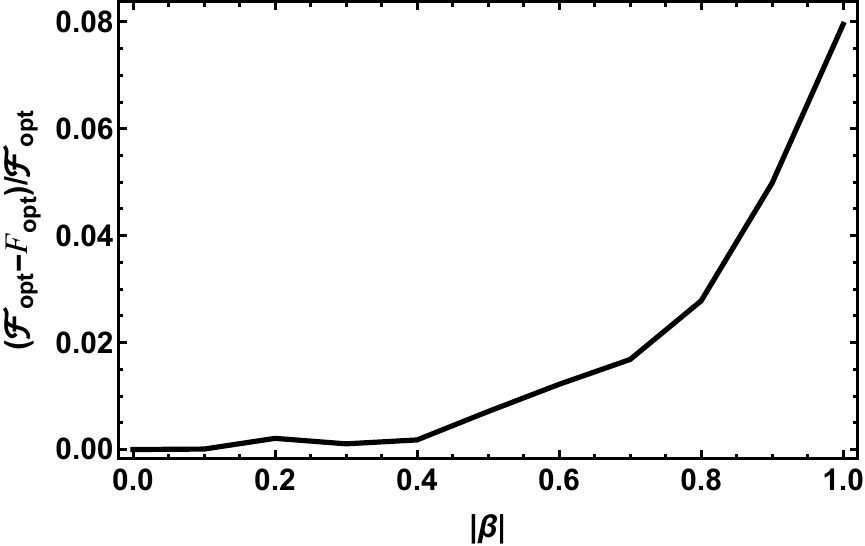}
    \justifying
    \caption{The relative gap between the QFI and the dyne-optimal CFI, with both quantities maximized over time, is plotted as a function of $|\beta|$. Here, $\mathcal F_{\text{opt}}~(F_{\text{opt}})$ denotes the QFI (dyne-optimal CFI) optimized with respect to the evolution time. The other parameter values are $\omega = 2.1$, $\gamma = 0.05$, $n = 0.1$, and $\alpha = i$ with $\beta = -i\beta_I$.}
    \label{fig:GaussianMeas}
\end{figure}

A key observation emerges when comparing the optimized CFI with the QFI. In particular,  the gap between the dyne-optimal CFI and the QFI, both maximized over time, grows with increasing $\beta$, as shown in Fig.~\ref{fig:GaussianMeas}. This observation suggests that while Gaussian measurements suffice to attain the QFI with $\beta = 0$, they become progressively suboptimal as squeezing increases. This behavior possibly suggests that  achieving the QFI in the presence of strong squeezing may require  non-Gaussian measurements, which go beyond the general-dyne framework including photon-counting and other nonlinear detection strategies.

%progressively higher degrees of non-Gaussianity in 
%the measurement may be required to achieve the QFI as $\beta$ increases. 
%Here as well, Fig.~\ref{fig:GaussianMeas} is shown for the parameter values $\omega = 2.1$, $\gamma = 0.05$, $n = 0.1$, and $\alpha = i$ with $\beta = -i\beta_I$; nevertheless, as in the previous case, the same conclusions extend to other parameter values within the feasible range. 
Therefore, our analysis also highlights the potential role of non-Gaussian resources in quantum metrology. We also emphasize here that although Figs.~\ref{fig:optdyne} and ~\ref{fig:GaussianMeas} are presented for specific parameter choices, the qualitative features discussed above remain robust across a wide range of physically relevant regimes.

%From Fig.~\ref{fig:dyne_Non_Markovian}, 

\begin{figure}
    \centering
    \includegraphics[width=1.0\linewidth]{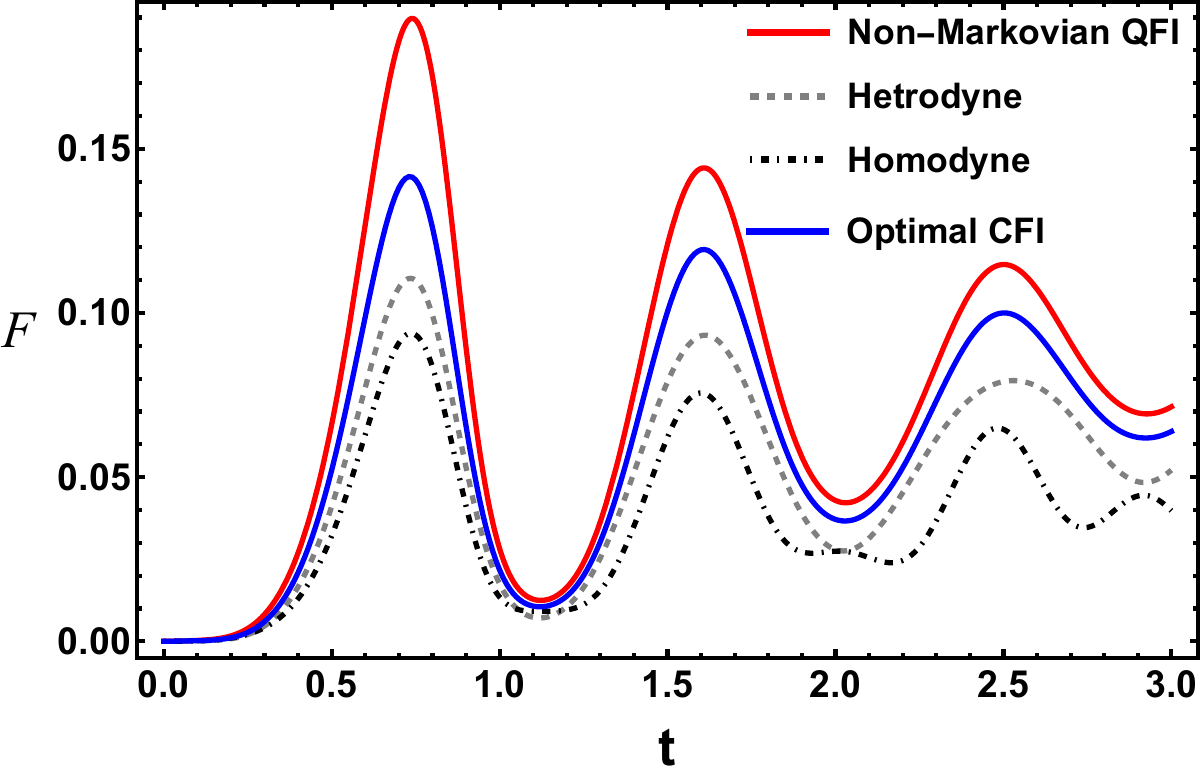}
    \caption{ CFI for different measurement strategies, homodyne, heterodyne, and optimal measurements, compared with the non-Markovian QFI. The other parameter values are $\omega = 7$, $\xi = 0.3$, and $\Lambda_{c} = 1$, $\alpha=0.1$, $k_{B}T/\hbar \Lambda_{c}=1000$.}
    \label{fig:dyne_Non_Markovian}
\end{figure}

When the system interacts with the environment described in the preceding section, we observe that optimal dyne measurement closely approaches the QFI over a significant time window (as depicted in Fig. \ref{fig:dyne_Non_Markovian}), confirming that Gaussian measurements remain nearly optimal even in the presence of environmental memory. However, both homodyne and heterodyne detection exhibit a noticeable gap from the QFI, particularly during periods where non-Markovian effects are dominant. This deviation in performance arises because information backflow dynamically reshapes the phase-space geometry of the evolved state and, consequently, changes the optimal measurement basis. Fixed quadrature measurements cannot adapt to this time-dependent optimal direction and therefore fail to fully extract the available information.

\section{Conclusion}
\label{sec:conclu}

Continuous variable (CV) quantum systems, especially Gaussian paradigm,  provides an experimentally accessible framework, with applications ranging from quantum computation, and quantum communication to quantum metrology. Understanding the ultimate precision limits of CV probes in realistic noisy environments is, therefore, of both fundamental and practical importance. 

%In this work, we have 

Our work examined the precision limits of frequency estimation in noisy CV system by combining the system engineering
%a dynamic control scheme 
with a detailed treatment of environmental effects. Our analysis reveals that dissipative processes impose a fundamental limit on achievable precision, leading to a finite optimal encoding time. Thus, unlike ideal unitary dynamics where information can accumulate indefinitely, the system-environment interaction creates a tradeoff between parameter encoding and decoherence 
%even in simple thermal environments. 
We demonstrated that introducing squeezing at the Hamiltonian level modifies the encoding process and provides a systematic route to enhance sensitivity, particularly in the early stages of the evolution. 
In particular, in the presence of a Markovian bath, we observed that squeezing-induced nonlinearities are capable of increasing quantum Fisher information (QFI), thereby establishing a dynamical control as an effective method for improving metrological performance. 
%where higher order time contributions become relevant.

Another important outcome of our study is the identification of non-Markovian dynamics as a beneficial feature in this setup. 
%rather than a detrimental one. 
Specifically, memory effects arising from structured environments enable partial recovery of information loss and can temporally restore the distinguishability of probe states.
%thereby enhancing estimation performance beyond what is achievable in a purely Markovian regime. 
{As a consequence, the QFI can exhibit transient revivals and precision enhancements beyond those attainable under unitary and Markovian dynamics. } These effects are especially pronounced in the
%transient 
intermediate-time regime, where the interplay between diffusion and dissipation leads to a non-trivial behavior of the QFI.
%In the last part , 

We also addressed the achievability of QFI  via experimentally implementable measurements. We found that the Gaussian detection schemes remain highly effective across a broad range of parameters, with optimized general-dyne measurements often approaching the ultimate bounds. %become more complex, particularly with
However, when the system dynamics incorporates stronger squeezing, the gap between the precision obtained through Gaussian measurements and QFI can increase, indicating that more general, potentially
%achieving the theoretical limits may require more general, potentially 
non-Gaussian, measurement strategies may be required to saturate the theoretical limit. 
Our results demonstrate that Hamiltonian design and reservoir engineering provide two complementary pathways to mitigate noise and enhance metrological performance in CV architectures. This framework offers a versatile platform for developing robust quantum sensing protocols in realistic open system scenarios, and may be directly realizable in several experimental setups, including optical and optomechanical architectures.

%\label{sec:conclu}

\acknowledgements
A.P., M., A.S.D. acknowledge the support from the project entitled ``Technology Vertical - Quantum Communication'' under the National Quantum Mission of the Department of Science and Technology (DST)  (Sanction Order No. DST/QTC/NQM/QComm/2024/2 (G)). M. acknowledges the support from the Qmet Tech Foundation under Qmet PARIMANA Post-Doctoral Fellowship program 2025, funded by the Department of Science and Technology (DST), GoI, under the National Quantum Mission (Sanction Order No. Qmet/2025-10/HRD/PARIMANA/SL/PDF/QMET-515752). MGAP acknowledges partial support by EU and MUR under the project PRIN22 
 G53D23001110006-RISQUE. MGAP thanks Subhash Chaturvedi for the questions raised during the 
 IQWT2025 conference in Varanasi, which led to the development of this work.
 
\appendix
\onecolumngrid
\section{Preliminaries on Gaussian quantum states}
\label{app:Gaussianframework}
Since we are interested in analyzing scenarios where the probe states are quantum states of the electromagnetic field, we focus on Gaussian states~\cite{Braunstein_RMP_2005,Weedbrook_RMP_2012,Adesso_2014,Serafini_2017}. These states play a crucial role in several quantum information processing tasks due to their experimental relevance, as demonstrated in a variety of optical platforms~\cite{Furusawa_1998,Julsgaard_2001,Yoshikawa_PRA_2007,Yukawa_PRA_2008,Ryuji_PRL_2011}. To set the stage, we briefly summarize the relevant concepts of the Gaussian framework. A multi-mode optical field can be modelled as a collection of bosonic modes. Let us consider a continuous variable (CV) quantum system consisting of $N$ bosonic modes described by annihilation operators \(\hat{a}_k\) with \(k=1,\dots N\), which satisfies standard bosonic commutation relations \([\hat{a}_k, \hat{a}_l^\dagger] = \delta_{kl}\). The corresponding quadrature operators are defined as
\begin{equation}
\hat{q}_k = \frac{1}{\sqrt{2}} (\hat{a}_k + \hat{a}_k^\dagger), \quad
\hat{p}_k = \frac{1}{i\sqrt{2}} (\hat{a}_k - \hat{a}_k^\dagger),
\end{equation}
satisfying \([\hat{q}_k, \hat{p}_l] = \iota \delta_{kl}\), where we have taken \(\hbar=1\). It is convenient to group the canonical operators into a phase-space vector,
\begin{equation}
\hat{\mathbf{X}} = (\hat{q}_1, \hat{p}_1, \ldots, \hat{q}_N, \hat{p}_N)^{T},
\end{equation}
with commutation relations
\begin{equation}
[\hat{\mathbf{X}}, \hat{\mathbf{X}}^{T}] = i \Omega,
\end{equation}
where \(\Omega = \bigoplus_{k=1}^N \begin{pmatrix} 0 & 1 \\ -1 & 0 \end{pmatrix}\) is the symplectic form. 
A generic Gaussian state \(\hat{\rho}\) is completely characterized by its vector of first moment \(\langle \hat{\mathbf{X}} \rangle\) and its covariance matrix \(\bsigma\), defined as
\begin{align}
    \langle \hat{\mathbf{X}} \rangle &= \mathrm{Tr}[\hat{\rho} \hat{\mathbf{X}}], \\
    \bsigma &= \mathrm{Tr}\left[\hat{\rho}\left\{\hat{\mathbf{X}}-\langle \hat{\mathbf{X}} \rangle,(\hat{\mathbf{X}}-\langle \hat{\mathbf{X}} \rangle)^{T}\right\}\right],    
\end{align}
where \(\{.,.\}\) represents the anti-commutator. The covariance matrix \(\bsigma\) is a \(2 N \times 2N\) real symmetric matrix, which is strictly positive and satisfy the uncertainty relation (with \(\iota = \sqrt{-1} \))
\begin{equation}
        \bsigma + \iota \Omega \geq 0,
\end{equation}
ensuring the positivity of the quantum state.
\section{Computation of QFI for the parameter $\omega$, with special cases}
\label{app:QFI_computation}
Here, we present the explicit calculation of QFI for frequency estimation in the different dynamical regimes considered in the Sec. \ref{sec:squeezingadvantage}. We will show how the general Gaussian formalism leads to closed form expressions in relevant limits. For single-parameter estimation within Gaussian framework, the QFI associated with the parameter $\omega$ is given by Eq.~\eqref{eq:qfi_general}. Therefore, the problem reduces to determining
\begin{align}
\bm{u}(t) &= \frac{\partial \avg{\bX(t)}}{\partial\omega}~\text{and} \\
\bSigma(t) &= \frac{\partial \bsigma(t)}{\partial\omega}.
\end{align}
Under Gaussianity-preserving dynamics, the first and second moments evolve according to
\begin{align}
\frac{d\avg{\bX}}{dt} &= \bA \avg{\bX}~\text{and} \label{eq:mean_evo} \\
\frac{d\bsigma}{dt} &= \bA\bsigma + \bsigma \bA^T + \bD, \label{eq:cov_evo}
\end{align}
whose formal solutions are
\begin{equation}
\avg{\bX (t)} = e^{\bA t} \avg{\bX(0)}, ~\text{and}
\label{eq:appXt}
\end{equation}
\begin{equation}
\bsigma(t) = e^{\bA t} \bsigma(0) e^{\bA^T t} + \int_0^t e^{\bA s} \bD e^{\bA^T s} ds.
\label{eq:appsigmat}
\end{equation}
The drift and diffusion matrices are given by
\begin{align}
\bA &= \bOmega \bG - \frac{\gamma}{2} \id~\text{and} \nonumber\\
\bD &=\gamma (2n+1) \id,
\label{eq:appAandD}
\end{align}
where 
\[
\bG = \begin{pmatrix}
\omega + 2\beta_1 & 2\beta_2 \\
2\beta_2 & \omega - 2\beta_1
\end{pmatrix}, \quad
\bOmega = \begin{pmatrix} 0 & 1 \\ -1 & 0 \end{pmatrix},
\]
with $\beta = \beta_1 + i\beta_2$. Differentiating the mean vector with respect to $\omega$ gives
\[
\bm{u}(t) = \frac{\partial}{\partial\omega}\left[e^{\bA t} \avg{\bX}(0)\right].
\]
Using the identity for the derivative of a matrix exponential,
\[
\frac{\partial}{\partial\omega} e^{\bA t} = \int_0^t e^{\bA(t-s)} \frac{\partial \bA}{\partial\omega} e^{\bA s} ds,
\]
along with the relation $\frac{\partial \bA}{\partial\omega} = \bOmega$ and the expression for $\avg{\bX}(t)$ as given in Eq.~\eqref{eq:appXt}, we arrive at
\begin{equation}
\bm{u}(t) = \int_0^t e^{\bA(t-s)} \bOmega \avg{\bX(s)} ds=\int_0^t e^{\bA(t-s)} \bOmega  e^{\bA s} \avg{\bX(0)} ds. 
\label{eq:u_solution}
\end{equation}
Next, differentiating the Lyapunov equation with respect to $\omega$, and employing the relations
$\frac{\partial \bA}{\partial \omega} = \bOmega$ and $\frac{\partial \bA^T}{\partial \omega} = -\bOmega$, we obtain the differential equation
\[
\frac{d\bSigma}{dt} = \bA\bSigma + \bSigma \bA^T + \bOmega \bsigma - \bsigma \bOmega,
\]
subject to the initial condition $\bSigma(0)=0$. The corresponding solution can then be expressed as
\begin{equation}
\bSigma(t) = \int_0^t e^{\bA(t-s)} [\bOmega \bsigma(s) - \bsigma(s) \bOmega] e^{\bA^T(t-s)} ds. 
\label{eq:Sigma_solution}
\end{equation}
Substituting the above expressions into the general QFI formula in Eq.~\eqref{eq:qfi_general} yields the complete formal expression for the QFI.

\subsection*{Case 1: Unitary evolution (no noise, no squeezing)}
This regime describes purely unitary evolution in the absence of squeezing in the system Hamiltonian, with all parameters set to zero except the frequency $\omega$, i.e., $\beta = \gamma = n = 0$. From Eq.~\eqref{eq:appAandD}, the drift and diffusion matrices reduce to
\begin{align*}
\bA &= \omega \bOmega,\qquad\bD = 0.
\end{align*}
The resulting dynamics correspond to a rotation in phase space, described by
\[
e^{\bA t} = \begin{pmatrix} \cos(\omega t) & \sin(\omega t) \\ -\sin(\omega t) & \cos(\omega t) \end{pmatrix}.
\]
For an initial coherent state characterized by the displacement vector $\avg{\bX(0)}=\left(\sqrt2 ~\text{Re}(\alpha),\sqrt2 ~\text{Im}(\alpha)\right)^T$ and the covariance matrix $\sigma(0)=\id$, using Eqs.~\eqref{eq:appXt} and \eqref{eq:appsigmat}, the time-evolved first and second moments are given by
\[
\avg{\bX}(t) = e^{A t} \avg{\bX(0)}, \quad \bsigma(t) = e^{\omega \bOmega t} e^{-\omega \bOmega t} = \id.
\]
Since the covariance matrix remains unchanged throughout the evolution, one immediately has $\bSigma(t)=0$. Furthermore, due to the commutativity between $\bA$ and $\bOmega$, the derivative of the mean vector as given in Eq.~\eqref{eq:u_solution} simplifies to
\[
\bm{u}(t)= t\,\bOmega \avg{\bX}(t).
\]
The QFI, therefore, becomes
\[
\mathcal{F}_\omega(t)=2t^2 |\avg{\bX}(t)|^2
=4t^2 |\alpha|^2,
\]
where we have used $|\avg{\bX}(t)|^2=2|\alpha|^2$, thereby demonstrating the quadratic (Heisenberg-like) scaling with time.
\subsection*{Case 2: Lossy evolution without thermal noise}
In this regime, the system evolves in the presence of a vacuum bath without any squeezing in the system Hamiltonian, i.e., $\beta=n=0$. Under these conditions, Eq.~\eqref{eq:appAandD} yields the drift and diffusion matrices in the form
\begin{align*}
\bA &= \omega \bOmega - \frac{\gamma}{2}\id, 
\qquad 
\bD = \frac{\gamma}{2}\id.
\end{align*}
Using Eq.~\eqref{eq:appXt}, the time evolution of the mean vector can be expressed as
\[
\avg{\bX}(t)=e^{-\gamma t/2} e^{\omega \bOmega t}\avg{\bX}(0).
\]
On the other hand, from Eq.~\eqref{eq:appsigmat}, the covariance matrix is found to remain invariant throughout the dynamics, i.e., $\bsigma(t)=\id$, which immediately leads to
\[
\bSigma(t)=\frac{\partial \bsigma(t)}{\partial\omega}=0.
\]
Furthermore, since $\bA$ commutes with $\bOmega$ in this regime, one obtains
\(
e^{\bA(t-s)} \bOmega e^{\bA s}
= \bOmega e^{\bA t}.
\)
Using this relation, the derivative of the mean vector as given in Eq.~\eqref{eq:u_solution} can be written as
\[
\bm{u}(t)
= t\,\bOmega e^{\bA t}\avg{\bX}(0)
= t\,\bOmega \avg{\bX}(t).
\]
Therefore, the QFI takes the form
\[
\mathcal{F}_\omega(t)
=2t^2
\big[\bOmega \avg{\bX}(t)\big]^\top
\big[\bOmega \avg{\bX}(t)\big]
=2t^2 |\avg{\bX}(t)|^2
=4t^2 |\alpha|^2 e^{-\gamma t}.
\]
It is worth noting that although the QFI initially exhibits a quadratic growth in the short-time regime, at later times it undergoes an exponential decay governed by the damping rate $\gamma$.
\subsection*{Case 3: Loss with thermal noise}
In this regime, the system undergoes evolution in contact with a thermal bath characterized by photon number $n$, while the system Hamiltonian contains no squeezing term, i.e., $\beta=0$. Under these assumptions, using Eq.~\eqref{eq:appAandD}, the drift and diffusion matrices take the form as
\begin{align*}
\bA &= \omega \bOmega - \frac{\gamma}{2}\id, \\
\bD &= \gamma\left(2n+1\right)\id.
\end{align*}
Similar to Case $2$, the evolution of the mean vector, as obtained from Eq.~\eqref{eq:appXt}, is given by
\[
\avg{\bX}(t)
= e^{-\gamma t/2} e^{\omega \bOmega t}\avg{\bX}(0).
\]
On the other hand, the covariance matrix satisfies the differential equation
\[
\frac{d\bsigma}{dt}=\bA\bsigma+\bsigma \bA^T+\gamma\left(2n+1\right)\id,
\]
for which the corresponding solution is found to be
\[
\bsigma(t)=\Big[1+2n(1-e^{-\gamma t})\Big]\id.
\]
Since $\bA$ commutes with $\bOmega$, the derivative of the mean vector again simplifies to
\[
\bm{u}(t)=t\,\bOmega \avg{\bX}(t).
\]
Furthermore, from Eq.~\eqref{eq:Sigma_solution}, one finds
\[
\bSigma(t)=\int_0^te^{\bA(t-s)}\left[\bOmega \bsigma(s)-\bsigma(s)\bOmega\right]e^{\bA^T(t-s)}\, ds=0,
\]
because $\bsigma(s)$ is proportional to the identity matrix and therefore commutes with $\bOmega$. As a result, the QFI takes the form as
\[
\mathcal{F}_\omega(t)=2t^2\big[\bOmega \avg{\bX}(t)\big]^\top\bsigma^{-1}(t)\big[\bOmega \avg{\bX}(t)\big]=\frac{2t^2 |\avg{\bX}(t)|^2}{1+2n(1-e^{-\gamma t})}.
\]
Using
\(
|\avg{\bX}(t)|^2=2|\alpha|^2 e^{-\gamma t},
\)
we finally obtain
\[
\mathcal{F}_\omega(t)=\frac{4t^2 |\alpha|^2 e^{-\gamma t}}{1+2n(1-e^{-\gamma t})}.
\]
\subsection*{Case 4: General Case}
We now turn to the most general scenario considered in this work, where the system evolves in the presence of a thermal bath with photon number $n$ together with squeezing in the system Hamiltonian. In this case, the dynamics is governed by Eq.~\eqref{eq:evolution}, with the Hamiltonian given in Eq.~\eqref{eq:Ham_withsqueezing}. Although no compact analytical expression for the QFI can be obtained in this regime, the problem can still be treated numerically. To this end, it is convenient to decompose the drift matrix as
\[
\bA = -\frac{\gamma}{2}\id + \bB,
\]
where
\[
\bB =
\begin{pmatrix}
2\beta_2 & \omega - 2\beta_1 \\
-(\omega + 2\beta_1) & -2\beta_2
\end{pmatrix}.
\]
Consequently, the matrix exponential takes the form
\[
e^{\bA t}=e^{-\gamma t/2} e^{\bB t}.
\]
Defining
\(
\Delta = 4|\beta|^2 - \omega^2,
\)
the exponential term $e^{\bB t}$ admits different forms depending on the sign of $\Delta$.
\subsubsection*{Case 1: $\omega^2 > 4|\beta|^2$ \; ($\Delta < 0$)}
Introducing
\(
\kappa = \sqrt{\omega^2 - 4|\beta|^2},
\)
the matrix exponential can be expressed as
\[
e^{\bB t}=\cos(\kappa t)\id+\frac{\sin(\kappa t)}{\kappa}\bB.
\]
\subsubsection*{Case 2: $\omega^2 < 4|\beta|^2$ \; ($\Delta > 0$)}
For
\(
\kappa = \sqrt{4|\beta|^2 - \omega^2},
\)
the exponential matrix takes the form
\[
e^{\bB t}=\cosh(\kappa t)\id+\frac{\sinh(\kappa t)}{\kappa}\bB.
\]
\subsubsection*{Case 3: $\omega^2 = 4|\beta|^2$ \; ($\Delta = 0$)}
In the critical case $\kappa=0$, the exponential simplifies to
\[
e^{\bB t}=\id + \bB t.
\]
These explicit expressions make it possible to numerically evaluate all matrix exponentials and integrals appearing in the QFI expression given in Eq.~\eqref{eq:qfi_general} for the general scenario.
\twocolumngrid
\bibliography{SPbib}
\end{document}